\documentclass[journal]{vgtc}                

\ifpdf
  \pdfoutput=1\relax                   
  \usepackage{graphicx}                
  \DeclareGraphicsExtensions{.pdf,.png,.jpg,.jpeg} 
\else
  \usepackage{graphicx}                
  \DeclareGraphicsExtensions{.eps}     
\fi%

\graphicspath{{figures/}{pictures/}{images/}{./}} 
\usepackage{listings}
\usepackage{microtype}                 
\PassOptionsToPackage{warn}{textcomp}  
\usepackage{textcomp}                  
\usepackage{mathptmx}                  
\usepackage{times}                     
\usepackage{cite}                      
\usepackage{tabu}                      
\usepackage{booktabs}                  
\usepackage{enumitem}

\usepackage[dvipsnames]{xcolor}

\definecolor{cbPurple}{RGB}{117,112,179}
\definecolor{cbYellow}{RGB}{230,171,2}

\onlineid{1312}

\vgtccategory{Research}

\vgtcpapertype{Applications}

\title{Traveler: Navigating Task Parallel Traces for Performance Analysis}

\author{Sayef Azad Sakin, Alex Bigelow, R. Tohid, Connor Scully-Allison, Carlos Scheidegger, Steven R. Brandt,\\ Christopher Taylor, Kevin A. Huck, Hartmut Kaiser, Katherine E. Isaacs}
\authorfooter{

\item
 Sayef Azad Sakin and Connor Scully-Allison are with the University of Arizona. E-mail: \{sayefsakin,cscullyallison\}@arizona.edu.
\item
 Alex Bigelow is with Stardog.
\item 
 R. Tohid, Steven R. Brandt, and Hartmut Kaiser are with the Louisiana State University. E-mail: mraste2@cct.lsu.edu, \{sbrandt, hkaiser\}@lsu.edu
\item 
 Carlos Scheidegger is with RStudio.
\item 
 Christopher Taylor is with Tactical Computing Labs.
\item
 Kevin A. Huck is with the University of Oregon. E-mail: khuck@cs.uoregon.edu
\item
 Katherine E. Isaacs is with the University of Utah. E-mail: kisaacs@sci.utah.edu.
}

\abstract{
Understanding the behavior of software in execution is a key step in identifying and fixing performance issues. This is especially important in high performance computing contexts where even minor performance tweaks can translate into large savings in terms of computational resource use. To aid performance analysis, developers may collect an {\em execution trace}—a chronological log of program activity during execution. As traces represent the full history, developers can discover a wide array of possibly previously unknown performance issues, making them an important artifact for exploratory performance analysis. However, interactive trace visualization is difficult due to issues of data size and complexity of meaning. Traces represent nanosecond-level events across many parallel processes, meaning the collected data is often large and difficult to explore. The rise of asynchronous task parallel programming paradigms complicates the relation between events and their probable cause. To address these challenges, we conduct a continuing design study in collaboration with high performance computing researchers. We develop diverse and hierarchical ways to navigate and represent execution trace data in support of their trace analysis tasks. Through an iterative design process, we developed {\em Traveler}, an integrated visualization platform for task parallel traces. Traveler provides multiple linked interfaces to help navigate trace data from multiple contexts. We evaluate the utility of Traveler through feedback from users and a case study, finding that integrating multiple modes of navigation in our design supported performance analysis tasks and led to the discovery of previously unknown behavior in a distributed array library.
}

\teaser{
  \centering
  \includegraphics[width=\linewidth]{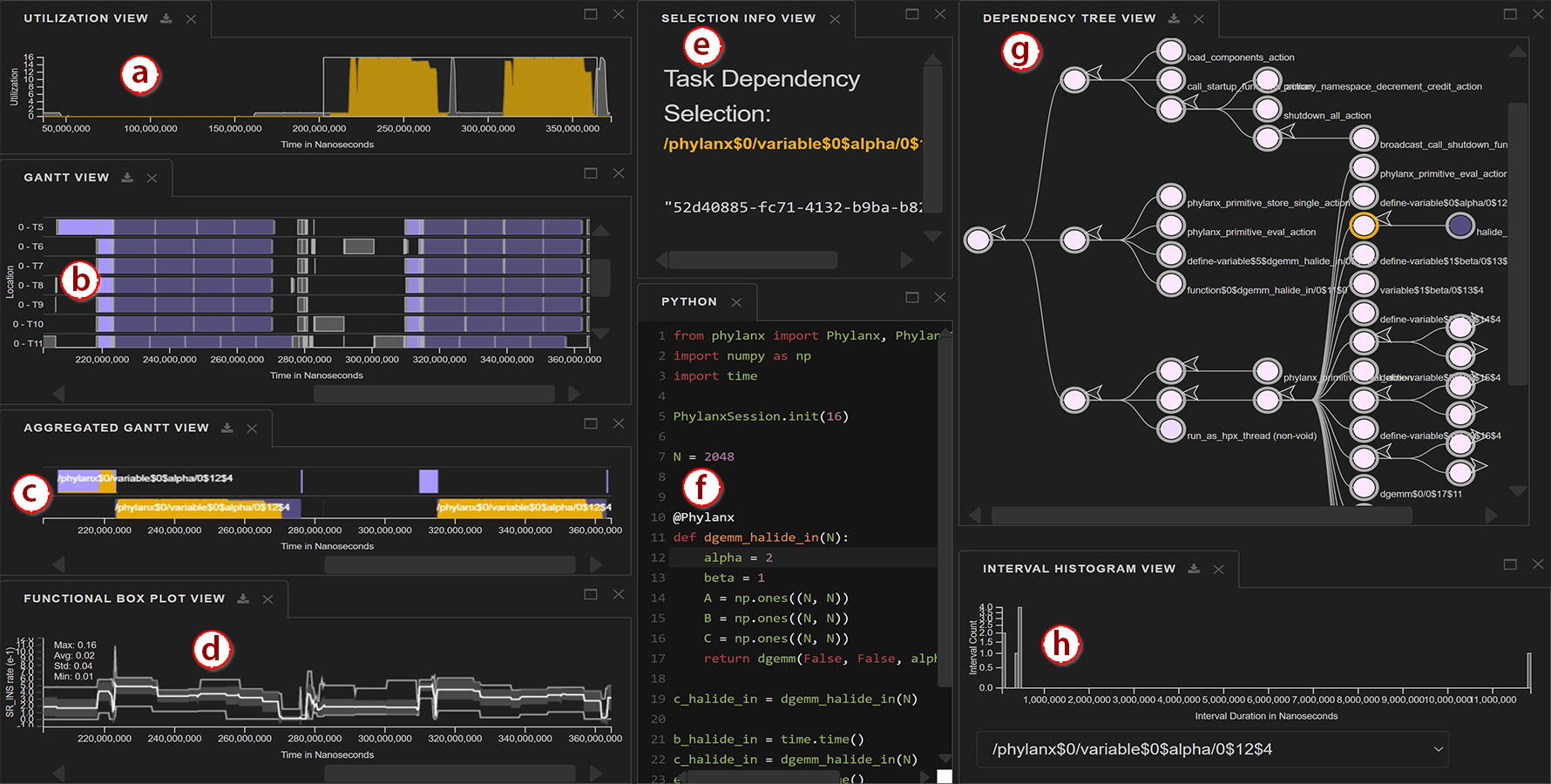}
  \caption{The Traveler interface. Traveler is a visualization platform supporting visual performance analysis of asynchronous task parallel programs. The Utilization View (a) shows the selected primitive (task type) accounts for most of the utilization in the program. The Gantt View (b) shows how individual tasks are scheduled across the system while the Aggregated Gantt View (c) shows utilization and extents due to groups of tasks. A Functional Box Plot View (d) summarizes performance metrics across resources and over time. Selection details are shown in the Selection Info View (e) and source code in the Source Code View (f). The Dependency Tree View (g) shows relations among task types. The distribution of task durations can be explored in the Interval Histogram View (h).
  }
  \label{fig:teaser}
}

\vgtcinsertpkg

\begin{document}

\firstsection{Introduction}

\maketitle

High performance computing (HPC) is essential in scientific simulation and discovery. HPC resources however are limited and costly. There is a great need to optimize the performance of programs running on them, as even a
slight performance improvement can lead to a significant reduction in  resources used~\cite{ashby2010opportunities}, thereby allowing more scientific applications to execute. 

Understanding the performance of these programs, a first step towards optimization, is difficult due to the complicated nature of underlying hardware components, intermediate execution scheduling systems, and the applications themselves~\cite{vetter2017contemporary}. One common approach to examining this behavior is to analyze an {\em execution trace}---a chronological record of per-resource events captured during program execution. 

Traces permit reconstruction of what occurred during execution, but gaining such insights at a useful level is challenging as the scale of events is orders of magnitude smaller than the length of the trace. Identifying behaviors of the program or system from the trace events that lead to them requires navigating a large space where details matter. 

Execution traces are typically visualized using a Gantt chart. To view details, users must examine very small time slices due to the short duration of trace events. Statistical overviews, such as stacked area charts showing the mix of executing items, help navigate to some degree, but the vastness of the trace still requires a lot of manual searching. We design Traveler, a multiple coordinated view system for execution traces, with multiple levels of meaningful abstraction for task parallel programs. Traveler links the expression tree of the program and provides aggregated Gantt views to help users relate trace events to code and show how individual iterations manifest on the system.

We designed Traveler through an iterative process with an active HPC development team. We followed the Design Study Methodology~\cite{Sedlmair2012} with the recommendations for evolving data and concerns of Williams et al.~\cite{williams2020movingtarget}. Through this process, we documented the changing needs of our collaborators over time and extended an existing task analysis for performance analysis. We discuss benefits and challenges of an ongoing, evolving design study in \autoref{sec:reflection}.

We validate Traveler through a case study, an interview with an expert user of several months, and feedback sessions with novice users. Our case study describes how Traveler was used to explain performance differences between two program implementations and to identify a performance bug, leading to a fix in the application's source code. 

The contributions of this study are as follows:
\begin{itemize}
    \itemsep=0.25ex
    \item A data and task analysis for performance analysis of task parallel traces with evolving concerns,
    \item The design and implementation of Traveler, a visualization system for performance analysis of task parallel traces, and
    \item Reflections and recommendations for ongoing design study collaborations and for visualization design of large execution traces. 
\end{itemize}

\section{Task Parallel Programs and Execution Traces} 
\label{sec:background}

Parallel programs aim to achieve faster computation by dividing work among many resources (e.g., CPUs, hardware threads) executing in parallel. We refer to concerns such as the time-to-solution and efficiency of resource use as {\em performance}. Good performance allows the scientific community to run larger or more numerous simulations and computations. In some cases performance improvements are required for a particular problem to be feasibly computed at all. Analyzing performance is difficult because ``good performance'' does not have a clear definition and there is a vast space of possible performance problems. It is unknown whether the observed execution time of a program is near optimal or longer than necessary due to a performance issue.

We consider a class of parallel programs that are managed by an {\em Asynchronous Tasking Runtime} (ATR). We call these programs {\em task parallel}. In the ATR paradigm, the program will break a problem into numerous {\em tasks}---units of computational work. Some tasks will generate other tasks as the problem is further divided. The assumption is that there will be many more tasks than resources. The ATR then schedules the tasks for execution on the parallel resources and may move tasks between these resources to better utilize them. Hence, one performance metric of interest for ATRs is {\em utilization}---the amount of parallel resources active (i.e., not {\em idle}) throughout the execution.

ATRs are different from {\em bulk synchronous} parallel applications, where resources run the same computation on different parts of the data, synchronize, and repeat the cycle. In a bulk synchronous paradigm, when the work required for each portion of the data is different, resources with less work may be idle while waiting for others. ATRs aim to take advantage of this waste by assigning the idling resources to other, potentially different work. In comparison to bulk synchronous applications, task parallel programs have smaller and more numerous tasks that may not be scheduled in a regular fashion due to a more complicated scheduling approach, making it difficult to determine what the system is doing and why at any given time.

{\em Execution Traces} are a form of performance data that logs temporal events and their {\em location}, i.e., the resource on which they occurred. In ATRs, task start time and end time are typically recorded as point events and then interpreted as a single durational event. The duration of these events is typically many orders of magnitude smaller than the entire length of the trace. In some execution traces, including ours, dependency data between tasks is also recorded. In our case, a dependency indicates a parent-child relationship, that one task caused another task to be created and ultimately executed. 

Execution traces can be seen as event sequence data. In terms of the survey by Guo et al.~\cite{guo2021survey}, traces are {\em high-dimensional}---task types are defined by their method/source code and thus there may be hundreds, if not more, different types; traces are (ideally) {\em dense}, at least when utilization is high; traces are {\em irregular} as tasks are asynchronously scheduled; and trace events {\em occur in parallel} across resources. Unlike prevalent forms of event sequence data, execution traces describe events generated by computer programs rather than human actors. Thus, rather than discovering frequent patterns of events, analysis tasks for execution traces are often more focused on identifying events, correlating them with code, and understanding how the state of the system and the code worked in tandem to achieve the observed performance. We discuss performance analysis tasks further in \autoref{sec:taskanalysis}.

\section{Related Work}
\label{sec:related}

Visualization has been widely-used for performance analysis and debugging in high performance computing~\cite{isaacs2014state, ezzati2017multi}. Among them, execution trace data is typically visualized with Gantt charts~\cite{nagel1996vampir, zaki1999toward, graham2004gprof, reinders2005vtune, adhianto2010hpctoolkit,Drebes2014Aftermath, zhukov2015scalasca,Pinto2016,xie2018visual}---a visual idiom where parallel timelines for each computing resource are arranged as rows and events, such as function calls, are drawn as rectangles within those rows, spanning their time of execution. Dependencies, if depicted, are represented as lines connecting events. 

Gantt charts are often paired with other statistical views. Overviews of event type or utilization are particularly common. SmartTraces~\cite{Osmari2014SmartTraces} implemented a drag-and-drop system of linked views, including a data-flow graph. Traveler similarly has a configurable system of auxiliary views. We add two structural views, an Aggregated Gantt View and a Dependency Tree View to show meaningful levels of abstraction between overview and detail.

Data scale has been a persistent issue in execution trace visualization. In addition to responsiveness issues, the vast difference in scale between the length of the trace and the individual events can make Gantt charts hard to interpret, especially when dependencies are present. SyncTrace~\cite{Karran2013SyncTrace} has a resource-centric approach where the timeline of one resource is shown in multiple levels of detail with the most zoomed in showing connections to other resources.  Ravel~\cite{isaacs2014combing} used an idealized unit time axis to show dependency patterns. However, this approach assumes a bulk synchronous paradigm that is not present in our data. Haugen et al.~\cite{Haugen2015} show dependencies only for a selected interval. We similarly provide dependencies on interaction, but show the entire chain of dependencies related to an event rather than just direct connections.

Several approaches eschew the Gantt chart and plot events or metrics about resources in the same space~\cite{muelder2009visual, muelder2016visual, fujiwara2018visual, li2019visual, kesavan2020visual}. We also aggregate metric data by resource in our metric views. However, these approaches do not preserve dependencies between the events, one of the reasons we choose to preserve the Gantt chart.

Non-timeline approaches use summaries, animation, or networks.
Sigovan et al.~\cite{sigovan2013visualizing} show events bubbling up a duration axis, using animation for time. Sanderson et al.~\cite{sanderson2018coupling} show streaming data for performance analysis and steering, with statistical plots and the projecting of metrics onto the 3D simulation space and machine room layout. Chuimbuko~\cite{kelly2020chimbuko} also shows streaming data, with statistical plots and a call stack plot of streaming data. Of these, only Chimbuko's call stack plot shows dependencies and like SyncTrace it is limited to a specific time range and focus resource. Our users have tasks requiring more fidelity in how resources, events, and dependencies work in concert.

As scheduling in task parallel programs is defined by an execution graph, per-event layered node-link diagrams have also been used~\cite{Huynh2015DAGViz,Muddukrishna2016GrainGraphs,Reissmann2017GrainGraphs}, but the simplification techniques to keep the graph scalable rely on a fork-join parallelism model that does not match our data.

\section{Our Design Process}
\label{sec:process}

We describe the design process of Traveler in terms of the Design Study Methodology of Sedlmair et al.~\cite{Sedlmair2012}. We first describe the context of the collaboration in which Traveler was designed, in support of a highly active open-source research software endeavor, making day-to-day needs a ``moving target''~\cite{williams2020movingtarget}. From this framing we describe roles in the project and our discovery process. Finally, we present our task analysis, which extends and modifies an existing goal-task lattice~\cite{williams2020movingtarget}.

\subsection{Collaboration Overview and Background}
\label{sec:collaboration}

We designed and developed Traveler in support of {\em and} as one of the goals of a multi-institutional research project, Phylanx. The overall goal of the project is to support distributed processing of data science programs, thus increasing the size of programs that can be computed and the speed in which they are computed. The Phylanx workflow transpiles Python/NumPy programs and runs them as distributed HPX applications. HPX is an asynchronous tasking runtime.

Performance analysis and visualization support are sub-goals of the Phylanx project supporting the development of both the core Phylanx library as well as HPX, application analysis by advanced Phylanx users, and communication of Phylanx concepts to potential users. 

As both Phylanx and HPX are actively developed, specific (visual) analysis goals of focus change over time, requiring careful observation of evolving needs in designing visualizations. Williams et al.~\cite{williams2020movingtarget} documented this process in the design of another performance visualization, {\em Atria}. While Traveler focuses on execution trace analysis, Atria was designed for profile data---a different type of performance analysis data where measurements are aggregated rather than presented in time. We discuss these implications of this difference in \autoref{sec:taskanalysis}. The assignment of general roles and the discovery process we use here are consistent with that of Atria.

{\bf Participants and Roles.} The Phylanx project is composed of three teams. The {\em Runtime Team} develops the HPX and the core Phylanx library. The {\em Performance Analysis Team} develops performance collection tools, automated performance optimization approaches, and nightly regression scripts. The {\em Visualization Team} develops performance visualization tools. This structure is unchanged from Atria's development, though the membership of each team has changed over time as people joined or left the project.

As with Atria, the Runtime Team lead served as the primary gatekeeper and encouraged the casting of students and post-doctoral fellows in their team as frontline analysts.  

\subsection{Data}
\label{sec:data}

We describe the data and data abstractions used in Traveler. 

{\bf Execution Trace Data.} Traveler visualizes execution trace data from Open Trace Format version 2 (OTF2)~\cite{OTF2} archives, a widely-used format in HPC. As previously stated, traces are a form of event sequence data. We abstract the fundamental unit of the trace (i.e., the durational event) as an {\em interval}. In ATR terms, an interval is similar to a {\em task}, but we avoid using the term ``task'' because HPX tasks can be further divided, so some intervals are not full tasks, and because the word ``task'' has a different meaning in visualization, one we use throughout the rest of the document.

An interval has a start and end time, a globally unique identifier ({\em GUID}), a {\em location} (computational resource, e.g., hardware thread) on which it executed, and a {\em primitive} name describing what was executed (e.g., function name or programming construct). The primitive is essentially the event {\em type}. An interval may also have a {\em Parent GUID}, the GUID of the interval that spawned it, if its parent was collected during execution. 

GUID and Parent GUID data in HPX were not collected prior to the Phylanx collaboration. These attributes were added to the trace data collection capabilities by the Performance Analysis Team lead after discussion with the Visualization and Runtime teams for some of the sub-goals we present in our task analysis, an example of evolving data that arose from the collaboration.  

The trace file may also include {\em performance counter} data. A performance counter accumulates a metric over time such as the number of CPU cycles or L1 cache misses. The value of this accumulator can be sampled at regular temporal intervals or at trace point events. Subtracting the counter value from its previously sampled one results in an absolute count of the events it measured in the past interval.

{\bf Execution Tree.} We derive an execution tree from the trace data, using the relationships described by Parent GUIDs. The execution tree aggregates the intervals by primitive name and sequence of primitives leading to it, similar to a calling context tree. We thus refer to each node as being a {\em primitive context}. Each node in the tree has an associated primitive name and aggregated duration of all intervals with the same context (i.e., sequence of parent primitives). Recursive calls may thus generate deep subtrees. The execution tree is different from Atria's execution graph which was collected directly rather than derived from the trace. Typically the derived execution tree is more detailed due to instrumentation differences.

{\bf Source Code.} Source code is an optional data source in Traveler.

\vspace{1ex}

\noindent{\bf Four Types of Data Entities.} The most fine-grained entity we associate data with is the {\underline interval}. However, performance analysts often think in terms of other entities that can be associated through the data: the computational {\underline resources}, functions in program ({\underline primitives}), or lines of source {\underline code}. We discuss these in our task analysis.

\subsection{Task Analysis}
\label{sec:taskanalysis}

Williams et al.~\cite{williams2020movingtarget} developed a task analysis with the goal of not only supporting Atria, but of informing performance data collection efforts in the other teams and preparing for future visual analysis needs as the Phylanx project continued. The task analysis was based on full team meetings, focused visualization or performance meetings, e-mail messages, and informal interviews. Through this process, the occurrence of tasks was tracked over time to prioritize tasks which remained relevant despite by the shifting needs of the project. We continued this discovery process in developing Traveler, adding an additional 181 note artifacts to the prior 152, and in turn continued to build and refine the task analysis. 

Two authors separately reviewed notes and proposed extensions to the existing task analysis and then discussed and integrated their proposals, resulting in the task analysis we present here. 

We originally organized tasks from broad umbrella concerns, to goals and sub-goals, and then finally low-level tasks, referring to the structure as a goal-task lattice as there were multiple connections between layers. For example, some goals support multiple umbrella concerns. Our revised goal-task lattice recognizes connections between layers that became apparent in our analysis. 

Another significant change was that we originally analyzed tasks for aggregated execution graph data only, using the network task taxonomy of Lee et al.~\cite{Lee2006}. However Traveler focuses on execution traces, a form of event sequences, but also incorporates an execution graph view, preserving those network-based tasks. We thus chose to apply the interaction task taxonomy of Yi et al.~\cite{Yi2007} for non-network tasks. We chose this task taxonomy due to the prevalence of {\em connect} tasks suggested by our sub-goals.

\begin{figure}[tb]
  \centering
  \includegraphics[width=\linewidth]{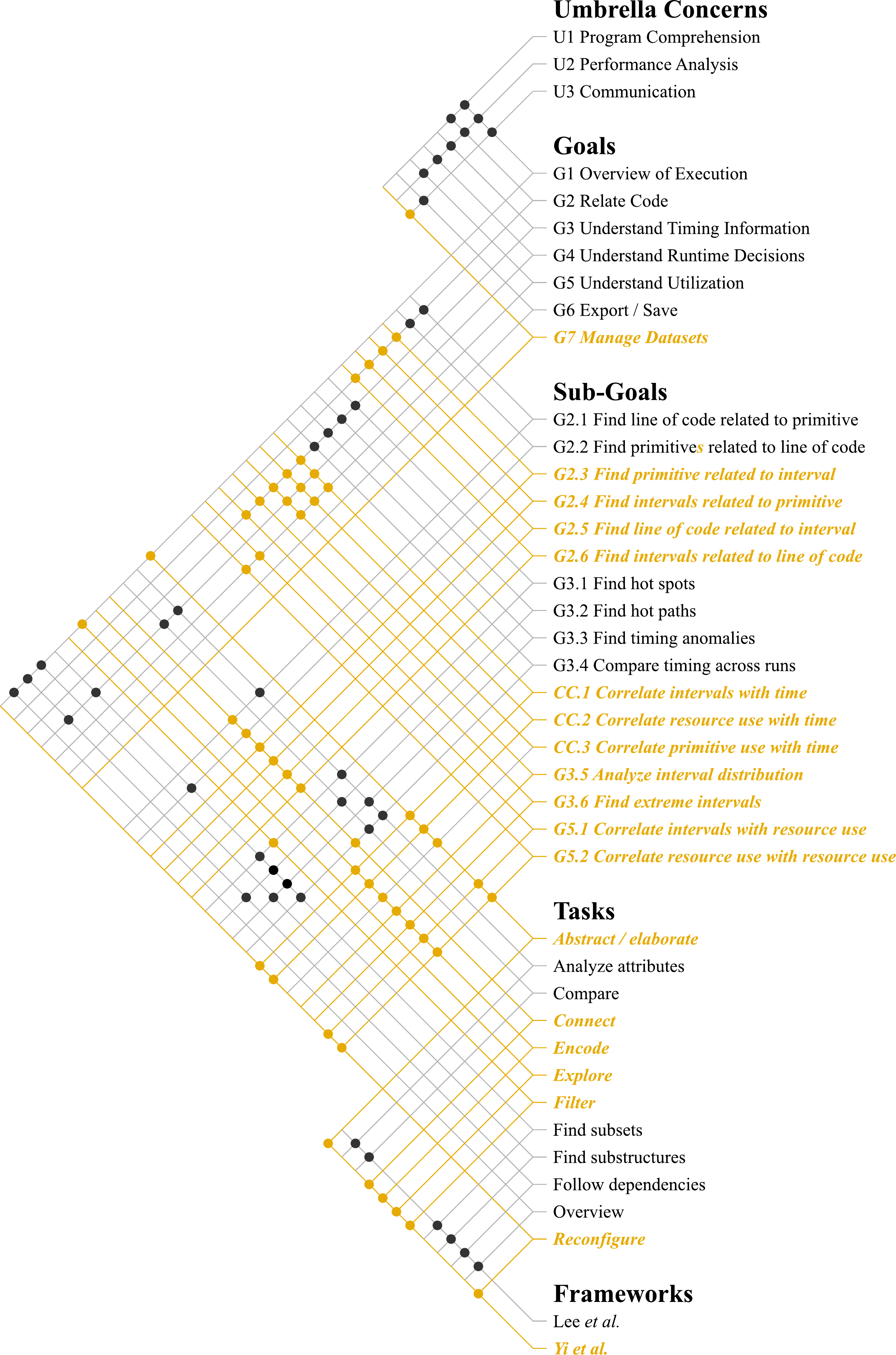}
  \caption{A goal-to-task lattice, showing relationships between high-level umbrella concerns, goals, sub-goals, tasks, and the task taxonomies in Lee et al.~\cite{Lee2006} and Yi et al.~\cite{Yi2007}. Additional goals, sub-goals, tasks, and frameworks, relative to Williams et al.~\cite{williams2020movingtarget}, are shown in \textbf{\textcolor{cbYellow}{gold}}.}
  \label{fig:taskanalysis}
 \end{figure}

We describe the unchanged parts of the task analysis briefly and elaborate on the changes. \autoref{fig:taskanalysis} provides an overview of the new analysis and its differences from the previous ones.

{\bf Umbrella concerns.} The three umbrella concerns did not change. {\bf U1}, {\em Program Comprehension}, describes how people want to understand and build mental models of how a program translates to execution. {\bf U2}, {\em Performance Analysis}, describes understanding efficiency concerns of the execution. {\bf U3}, {\em Communication}, describes concerns in explaining the system and approach to others, such as potential users or the academic community through publications.

{\bf Goals.} The six original goals did not change, but four of them were designated additional sub-goals and a seventh overall goal was added. We describe each goals and its sub-goals and tasks, with focus on the additions for execution trace data.

{\bf G1: Overview of Execution.} In all three umbrella concerns, people wanted an overview of the execution---a general sense of how the code was translated into parallel tasks, which were run frequently, and how time and resources were allocated. With the execution graph, this can be viewed with overview, follow dependencies, and finding substructures tasks. We add abstract and explore tasks from Yi et al. with the inclusion of trace data.

{\bf G2: Relate Code.} Understanding how behavior relates to code is important because the code is what the developers can change. Runtime developers need to know relationships to code to improve runtime policies. This is thus an important goal for program comprehension and performance analysis. In Atria, the sub-goals were relating primitive to code (analyze attributes) and code to primitives (find subsets). Intervals are the fundamental unit in trace data, so we add sub-goals that also relate intervals. These, as well as the previous sub-goals, are described by Yi et al.'s {\em connect} task.

{\bf G3: Understanding Timing Information.} Timing information is at the core of performance analysis. People want to know what took a significant or an abnormal amount of time and also want to compare timings across iterations or runs. Thus, we keep previously declared sub-goals of finding hot spots, hot paths, timing anomalies, and comparing across runs. 

The detailed temporal data about intervals and resources available in execution traces led our collaborators to discuss sub-goals focusing on them, three of which cross-cut several goals. The {\bf cross-cutting sub-goals} are (a) correlating intervals with time, (b) correlating resources with time, and (c) correlating primitive use with time. People want to know the arrangement of intervals and resources in time and at a higher level of abstraction: the primitive use in time. As the time span is vast in comparison to interval durations, several levels of detail are needed, leading us to assign abstract/elaborate, reconfigure, and encode tasks.

We identified two more timing information sub-goals which are highly related: analyze the interval distribution and find extreme intervals. Developers and performance analysts have questions regarding the duration and distribution of intervals. Long intervals may indicate an anomaly or target of optimization depending on numeracy. Many short intervals can also lead to poor performance due to the overhead in managing them. Understanding these in a timeline is unmanageable, suggesting an encode task.

{\bf G4: Understand Runtime Decisions.} An ATR manages parallel tasks to achieve good performance. Understanding the decisions made during execution is thus important for developers to best use the ATR and for Runtime Team members to improve the ATR. Our original task analysis recognized several network tasks. With execution trace data, the cross-cutting sub-goals also explain runtime decisions as the assignment of intervals and resources is managed by the runtime.

{\bf G5: Understand Utilization.} Utilization describes how computational resources are exercised during execution. High utilization means most of the resources are working at capacity and thus likely will achieve the result more quickly. Low utilization means many resources are idle, wasting power without contributing to the result. In our prior study, utilization was identified as a goal, but not one the data yet permitted. All utilization sub-goals are thus newly identified.

The cross-cutting sub-goals are central to utilization as refer to state information about resources in time. Additional sub-goals are (a) correlating intervals with resource use, explaining {\em what} computation is occurring if a resource is in use, and (b) correlating resource use with resource use---comparing across resources. Again as the time scale is large, abstract/elaborate and arrange tasks are necessary. Some questions, like other attributes of resources, require an encode task.

Goals G3, G4, and G5 influence each other, leading to the cross-cutting sub-goals. Runtime decisions affect the utilization, but utilization informs runtime decisions. Similarly, timing information is an outcome of runtime decisions, but like utilization, also informs them. This relationship became more apparent to us as the Utilization goal was expanded upon.  

{\bf G6: Export/Save.} Sharing of results supports the Communication concern. This is unchanged from the original task analysis.

{\bf G7 (New): Manage Datasets.} With more datasets as well as regression analyses and comparisons across versions or processor counts, our collaborators started discussing the need to manage multiple datasets during their analyses. We focus on execution trace visualization in this work and thus do not further discuss this goal.

\section{Traveler}
\label{sec:traveler}

We present the design of Traveler (\autoref{fig:teaser}), a visualization platform for performance analysis of task parallel execution traces. In \autoref{sec:travelerstrategy}, we provide an overview of Traveler's views, explaining how they match with the four data items types in our data (intervals, primitives, resources, and code) and how we build a hierarchy of levels of abstraction that are meaningful in the trace. We then discuss designs of the major views, temporal (\autoref{subsec:temporal_views}) and non-temporal (\autoref{subsec:non_temporal}), and how they fulfill our data and task analysis. Finally we discuss our implementation (\autoref{subsec:implementation}) and strategies we used to keep the visualization responsive.

\subsection{An Overview of Traveler Views}
\label{sec:travelerstrategy}

In support of the data and numerous tasks discussed in \autoref{sec:process}, Traveler is a configurable multiple coordinated view system. Users can re-size, arrange, hide, and close views as needed for their analysis goals. 

There are nine types of views in Traveler, some of which the user may place multiple times with different facets of the data. Of the views, we consider three to be the native views of the four data item types in execution traces. Intervals and resource state are shown together in a Gantt View, a widely-used idiom for trace data (see \autoref{sec:related}) that leverages user familiarity. Primitives are shown with their context (i.e., creation provenance) in a node-link diagram showing the program's execution tree, following the idiom used in Atria~\cite{williams2020movingtarget}. Source code is shown in its native text format (\autoref{fig:teaser}(f)) with syntax highlighting. 

Complimenting the Gantt chart, which can show individual event details, we provide two other temporal views at higher levels of abstraction that are meaningful to performance analysis: a utilization view and an aggregated Gantt chart which summarizes the behavior of meaningful groups of tasks. 

Traveler has two additional views for showing summarized interval or resource attributes (metrics) in time, a different facet of the intervals and resources. A histogram view of interval duration provides another selection and navigation mechanism into the trace data. 

Selections in the non-temporal views highlight the corresponding intervals in the Gantt View and show their distribution in time on the Utilization View. Analysts can then use the Utilization View to coarsely navigate to a time-span of interest and then use pan and zooming features to fine-tune the Gantt View.

The Selection Info View (\autoref{fig:teaser}(e)) shows details such as additional attribute data for any set of selected items from the other views.

\subsection{Temporal views}
\label{subsec:temporal_views}

Traveler has five views that use time for the horizontal axis, a Gantt chart, the novel Aggregated Gantt chart, a Utilization View, and two metric views: a simple Line Chart and the Functional Box Plot View. The Utilization View is fixed to the full duration of the execution trace. All other temporal views have linked panning and zooming as well as linked random access through Utilization View brushing.

\vspace{1ex}

\noindent
The \textbf{Utilization View} (\autoref{fig:teaser}(a)) shows the total utilization over time using an area chart. The height of the area chart at any pixel is the total amount of resource activity in that time span divided by the total time span, such that the maximum would be the total number of resources. This view supports G5 (understanding utilization) and at a summary level, cross-cut sub-goal CC.2, correlating resource use with time. 

The Utilization View also acts as a navigation aid and provides context to the other views. An interactive rectangular brush shows and can alter the time span in the other temporal views, supporting abstract/elaborate and reconfigure tasks. Primitive or group interval selections from other views will draw a second area in yellow, denoting the utilization due to the selection. This feature supports cross-cut sub-goals CC1 and CC3, correlating intervals and primitives with time.

\begin{figure}[h!]
    \centering
    \includegraphics[width=3.5in]{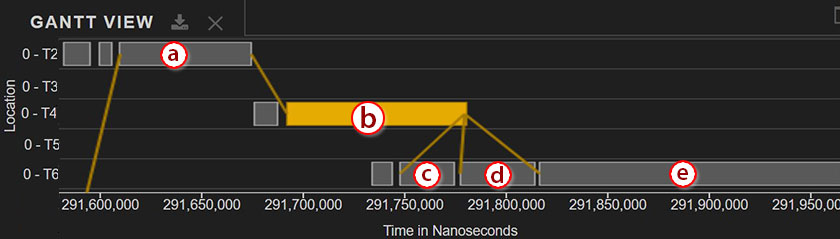}
    \caption{Gantt View representing parent-child dependency between two intervals. Here, the highlighted yellow interval bar $b$ has parent $a$. $b$ has three children\textemdash $c$, $d$, and $e$.
    }
    \label{fig:gantt}
\end{figure}

\vspace{1ex}

\noindent
The \textbf{Gantt View} (\autoref{fig:teaser}(b)) represents resource state over time as rows and intervals in time as rectangular bars on the resources on which they executed. This view supports cross cutting sub-goals CC.1 and CC.2, correlating intervals and resources with time, as well as G5.1, correlating intervals with resources. We label resources by their CPU and thread ID as $core\ ID-thread\ ID$. In addition to panning and zooming in time, users can pan and zoom over the space of resources. 

Users can select an individual interval on click, which will highlight the interval in yellow and update the Selection Info View and Utilization View. Additionally, parent-child relationships related to the selected interval will be shown using yellow lines drawn between the bars (\autoref{fig:gantt}). By showing these relationships,  users can pan and zoom to follow hot paths (G3.2). The relationships also give them insight into understanding runtime decisions (G4). 

We chose to show parent-child relationships on-demand-only because showing all of them is infeasible to interpret. While intervals of interest can be selected for further examination through visual inspection, the non-temporal views that help pick primitives by name, context, or duration help navigate to potential targets of investigation.

As the Gantt View is the native display of the {\em intervals}, it makes the details of individual events and dependencies available, but cannot summarize their behavior. The Utilization View on the other hand is too high-level to see all but the largest collections of events. We thus add a view in between that summarizes smaller groups of related intervals: the Aggregated Gantt View.

\begin{figure}[h!]
    \centering
    \includegraphics[width=\columnwidth]{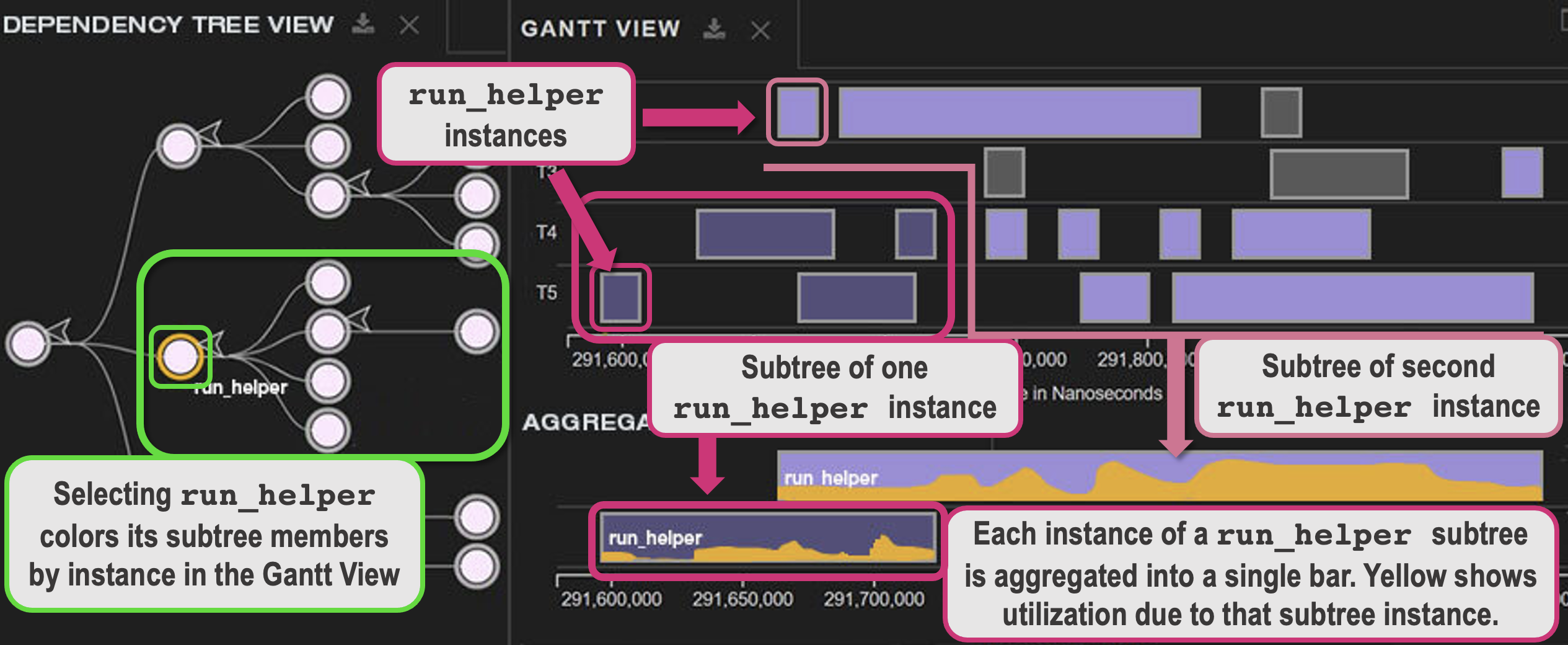}
    \caption{Selecting a primitive in the Dependency Tree View will highlight instances of that primitive's subtree in the Gantt View and create aggregated bars for the subtree in the Aggregated Gantt View.
    }
    \label{fig:agg_gantt}
\end{figure}

\vspace{1ex}

\noindent
The \textbf{Aggregated Gantt View} (\autoref{fig:teaser}(c)) provides a temporal layout to meaningful groups of intervals, as defined by the execution tree of the program. It is populated using the Dependency Tree View (\autoref{subsec:non_temporal}). The Aggregated Gantt View provides a simplified view at a higher level of abstraction than individual intervals of the Gantt View but a more detailed level of abstraction than the aggregated primitives of the execution tree, further supporting the correlation of primitives in time (cross-cut sub-goal CC3).

Each bar in the Aggregated Gantt View represents a specific instance of a primitive created by a sequence of dependencies as shown in \autoref{fig:agg_gantt}. For example, consider a primitive {\tt block} of code within a {\tt for} loop. Though there are several {\tt block}s throughout the program, the chain of dependencies leading to a particular {\tt block} uniquely defines it in comparison to the others. Because it is in a {\tt for} loop, it runs $k$ times. If shown in the Aggregated Gantt View, the {\tt block} will generate $k$ bars, showing where each was executed in time. 

Rather than showing only the {\tt block} interval, new aggregated intervals are created from all the intervals originally spawned by that particular {\tt block} instance, in other words, any interval the {\tt block} was an ancestor of. The horizontal position and length of each bar shows the starting time of its first interval and the ending time of its last. Because the different intervals spawned by the {\tt block} instance may occur on different resources, we cannot assign it to a single resource. Instead, we lay out the bars in a greedy fashion to avoid collision. Thus, the vertical position shows how many of these bars (e.g., loop iterations) run concurrently. 

Within each aggregated bar, we draw a yellow area chart showing the utilization due to that instance of the subtree. The background of each bar is shaded in purple with the subsumed intervals in the main Gantt View colored the same shade, aiding users in correlating intervals with primitives (sub-goal G2.4).

\vspace{1ex}

\noindent\textbf{Functional Box Plot and Line Chart Views.}
As explained in \autoref{sec:data}, performance counter (metric) data can be collected along with the trace in one of two ways: sampled with the interval events or sampled in time. Performance counters describe some attribute of the resource on which they are sampled. Traveler offers two views to show these metrics in time, supporting cross-cut sub-goal CC.2, correlating resource use with time. Furthermore, since these views are linked with the Gantt View and multiple of each of these views may be opened with different performance counters, they also support correlating resource use with resource use, sub-goal G5.2.

As performance counter data may be sampled irregularly, they are more meaningfully represented as a rate. For a consecutive $(time_1, value_1)$ and $(time_2, value_2)$ pairs where $time_1 > time_2$, the rate is calculated as, $rate=\frac{value_1-value_2}{time_1-time_2}$. We use this derived rate in our performance counter views.

The Functional Box Plot view summarizes performance counter rates across resources. We draw three lines representing the maximum value across resources, the minimum, and the average. A shaded gray area around the average shows a standard deviation. \autoref{fig:teaser}(d) shows the Functional Box Plot view for the  PAPI (Performance Application Programming Interface)~\cite{browne2000portable} metric CPU cycles. 

\subsection{Non-temporal views}
\label{subsec:non_temporal}

Traveler has four non-temporal views: the Dependency Tree (execution tree) View, an Interval Histogram View, a Source Code View, and a Selection Info View. We have previously described the Source Code View and Selection Info View. We discuss the other two views below.

\vspace{1ex}

\noindent
The \textbf{Dependency Tree View} (\autoref{fig:teaser}(g)) is the native view for primitives. We leverage this knowledge and design from Atria. It is a node-link tree visualization where every node represents a distinct primitive call as defined by its dependencies (e.g., the yellow lines in the Gantt View). As an example, a {\tt while} primitive may be generated from multiple places in the source code. By defining the sequence of tasks that generated a particular {\tt while} interval, we can group all {\tt while} primitives associated with the same overall action in the code. This is similar to a calling context tree in serial or synchronous programs. The difference is that the parent node---the primitive that created the child node---need not wait for its children to complete. 

The Dependency Tree View is also an expression tree describing how the source code is transformed into task types with dependencies which are then scheduled for execution. Each node in the Dependency Tree View represents numerous intervals all with the same primitive type and context. As we build the Dependency Tree View from the interval dependencies found in the trace, we can also derive aggregate metrics for the total time attributable to each node. We use a purple value ramp to indicate this aggregated time value, supporting subgoals G3.1-G3.4 in the same way Atria did.

Selecting a node in the Dependency Tree View will highlight the corresponding intervals in the Gantt View, draw a highlighted area on the Utilization View showing utilization due to the node, and generate the corresponding bars in the Aggregated Gantt View. This linking supports the cross-cutting subgoal CC3, correlating primitives with time. The link to the Gantt chart in particular also supports subgoal G2.4, relating primitives to intervals.

As the execution tree can get quite large, we only draw out the first five levels by default. Users can collapse or expand subtrees by clicking on the arrow in the upper right of each node. To reduce clutter, we do not label every node. Primitive names are available on hover.

\vspace{1ex}

\begin{figure*}[htp]
    \centering
    \includegraphics[width=0.49\textwidth]{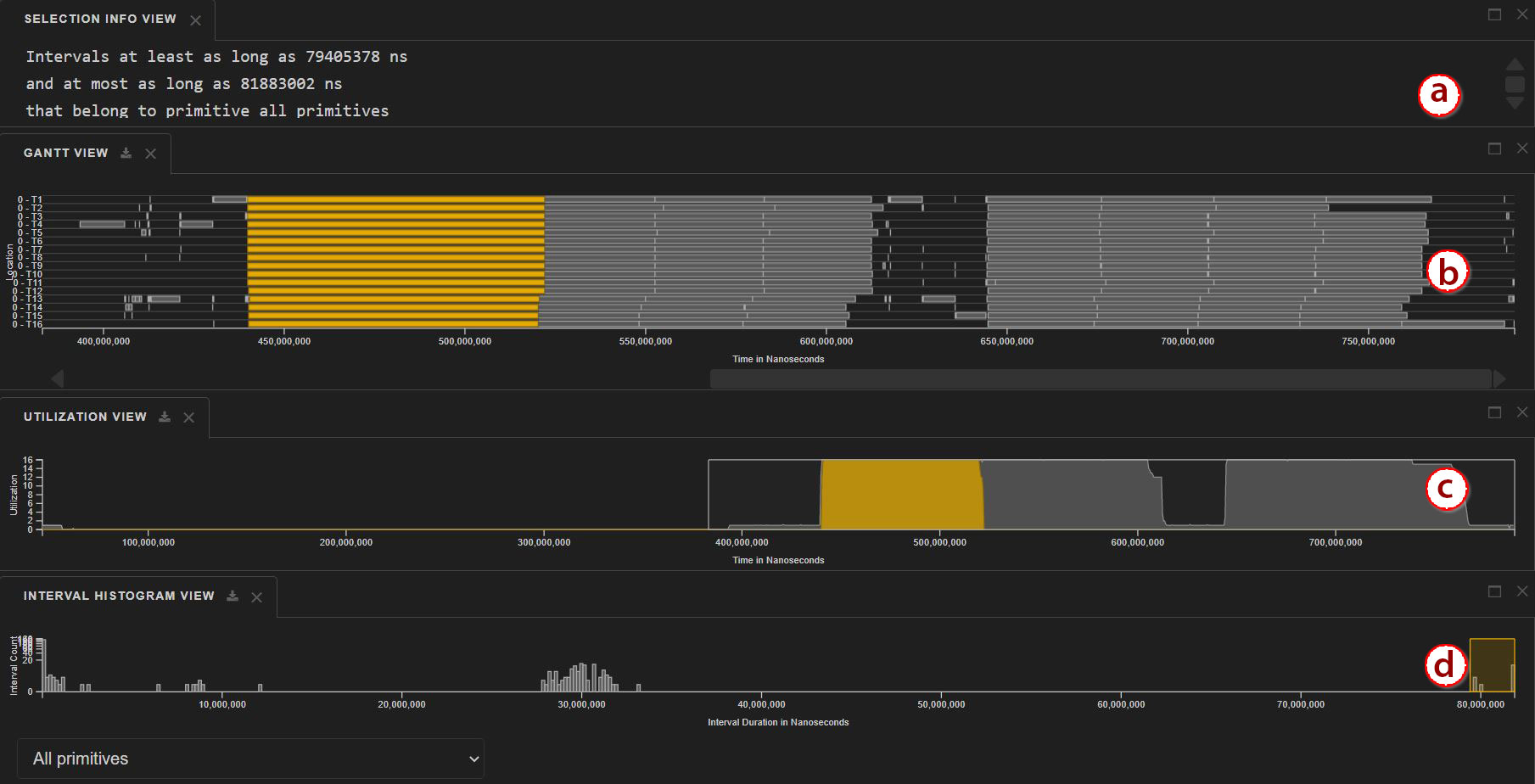}
    \hspace{1.5ex}
    \includegraphics[width=0.48\textwidth]{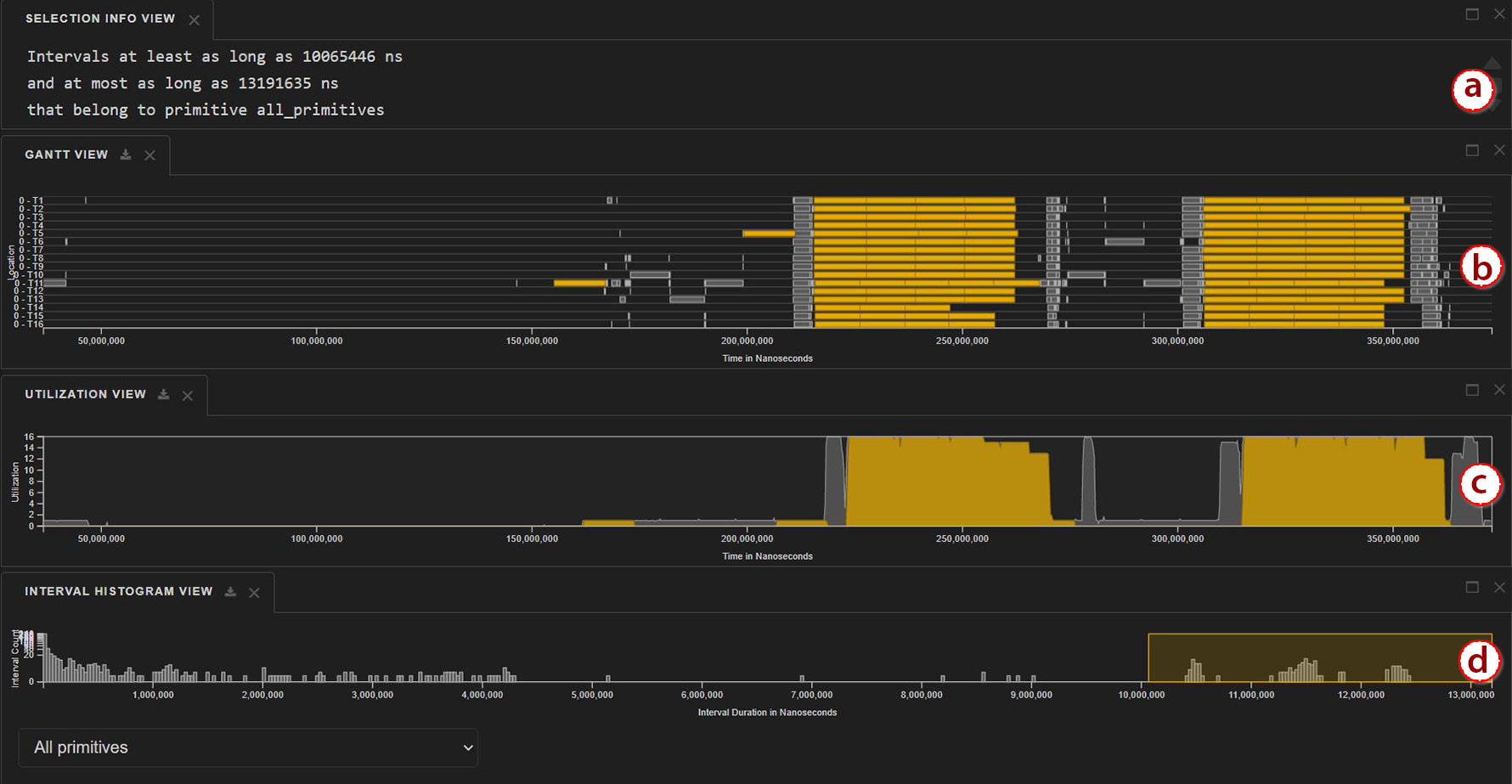}
    \caption{The distribution of intervals in Phylanx-Blaze (left) and Phylanx-Halide (right). The Selection Info View (a) shows the exact duration of the selected bars while the Gantt View (b) shows where they occur in time and on what resources. The Utilization View (c) shows their impact compared to the whole program, which is much higher in Phylanx-Blaze. The Interval Histogram View (d) shows the longest intervals in both have been selected via brush.
    }
    \label{fig:casestudy1}
\end{figure*}

\noindent
The \textbf{Interval Histogram View} (\autoref{fig:teaser}(h)) shows intervals binned by their duration. By default, all intervals are shown. A dropdown menu enables filtering to a particular primitive context as in the Dependency Tree view. Bins in the histogram can be selected via brushing. When selected, all intervals contributing to a bin will be highlighted in the Gantt View and Utilization View. This view directly supports the sub-goals G3.5 and G3.6, analyzing the interval distribution and finding extreme intervals. Outliers appear as isolated clusters of bars on far ends of the histogram, as shown in \autoref{fig:casestudy1}(d).

\subsection{Implementation}
\label{subsec:implementation}

Traveler is a web-based client-server application. It uses a Uvicorn server that communicates with the front-end using {\tt fastapi} in a RESTful manner. The back-end pre-processes the OTF2 file, building all data structures needed for data look up and fast rendering and storing them using DiskCache~\cite{diskcache}. Additionally, the back-end renders the primary drawing space of all temporal views as well as the Interval Histogram View for displaying on the front end using HTML5 Canvas. Axes and brushes are implemented on the front-end in Javascript using D3~\cite{bostock2011d3}. We provide a system diagram in the supplemental material.

\vspace{1ex}

\noindent\textbf{Data Structures.} We used {\em summed area tables}~\cite{summedarea} to enable fast generation of the temporal and histogram views and we used an {\em interval tree} for fast lookup of data attributes for the Selection Info View. These data structures are calculated at the time the data is initially given to Traveler, in a process we call {\em bundling}. Bundled datasets are saved by Traveler so the pre-processing step only occurs once. 

Specifically we pre-calculate summed area tables for CPU utilization, hardware counters, and duration count for each of the thread-core location and primitives, both individually and combined. When the front-end sends a request with the number of pixels available and the time range selected, we use the relevant summed area tables to generate a bitmap of pixel values which are rendered on the front end using HTML canvas. Retrieving new bitmap information induces latency, so we overdraw to hide this overhead when the user pans the view.

\section{Evaluation}
\label{sec:evaluation}

We validate Traveler through a case study which identified a previously unknown performance bug and resulted in changes to our collaborator's code. This study was done collaboratively between one member of the Runtime Team and one member of the Visualization Team and verified by the Runtime Team PI.

We then provide feedback from a non-author member of the Runtime Team who used Traveler over the course of several months. We also discuss feedback from novice Runtime Team members that we gathered during hour-long evaluation sessions on a cloud deployment.

\subsection{Case Study}
\label{subsec:casestudy}

We present a case study showing the utility of Traveler in analyzing performance in a real scenario. The work was collaboration between a Runtime Team member (RTM, a post-doc with several years of Phylanx experience) and a Visualization Team member (VTM, a graduate student) trying to understand the differences in performance between different implementations of Phylanx for a line of research the RTM was leading. The analysis took place over the course of four months interspersed with other priorities and activities for both team members, as performance analysis is only one part of the domain expert's activities.

We show how Traveler assisted in identifying and comparing computational resource utilization among multiple executions and helped reveal and ultimately fix a performance bug.

\vspace{1ex}

\noindent
\textbf{Analysis Problem and Data.}
The Runtime Team member is investigating potential benefits of using {\em Halide} in Phylanx in comparison to {\em Blaze}. {\em Halide}~\cite{ragan2017halide} is a framework for automatically synthesizing code for multi-platform systems with highly customizable task scheduling and data management features. {\em Blaze}~\cite{Blaze} is an open-source C++ library supporting high performance parallel linear algebra. We refer to the Phylanx using Halide as Phylanx-Halide and the Phylanx using Blaze as Phylanx-Blaze.

In exploring the differences between Phylanx-Halide and Phylanx-Blaze, the RTM used {\tt dgemm}~\cite{dgemm}, a well-known matrix-matrix kernel, as an example application because it exercises matrix multiplication, scaling, and addition. Initial performance results showed that the Phylanx-Halide version was nearly twice as fast as the Phylanx-Blaze version. To try to understand why, the RTM collected execution traces. The VTM used Traveler to investigate these traces and then shared findings with the RTM via text, screenshots, and movies.

\vspace{1ex}

\noindent\textbf{Investigation with Traveler.}
Viewing both execution traces in Traveler, we opened the Interval Histogram View (\autoref{fig:casestudy1}) to compare (Goal G3.4) the distributions in interval durations (G3.5). The Phylanx-Blaze distribution is much wider, with many intervals longer than the longest ones in Phylanx-Halide, indicating some tasks are taking much longer in Phylanx-Blaze. We select the long intervals in both traces (G3.6). The highlighted intervals in the Gantt show us the tasks are shorter and more evenly distributed in Phylanx-Halide. The Selection Info View reveals the extreme Phylanx-Blaze interval is six times longer than the longest one in Phylanx-Halide, contributing to the longer execution time. {\em Insight:} Phylanx-Blaze intervals require more time for the same work and are more unbalanced.

The RTM suspects that there may be data movement differences because data management is a feature of Halide. We open multiple Functional Box Plot views to check the L1 and L2 Cache miss rates (Sub-goal CC.2). (Figures are available in the supplemental material.) A higher miss rate suggests more time is being spent reading from slower memory. We brush over the Utilization View to focus on the second area of high activity in the computation. We see that while the fluctuations in L1 cache misses are similar, the Phylanx-Blaze version has a steadily increasing rate of L2 cache misses while the Phylanx-Halide version remains flat. {\em Insight:} the Phylanx-Halide version is managing data more effectively as suspected.

\begin{figure*}[h!]
    \centering
    \includegraphics[width=0.49\textwidth]{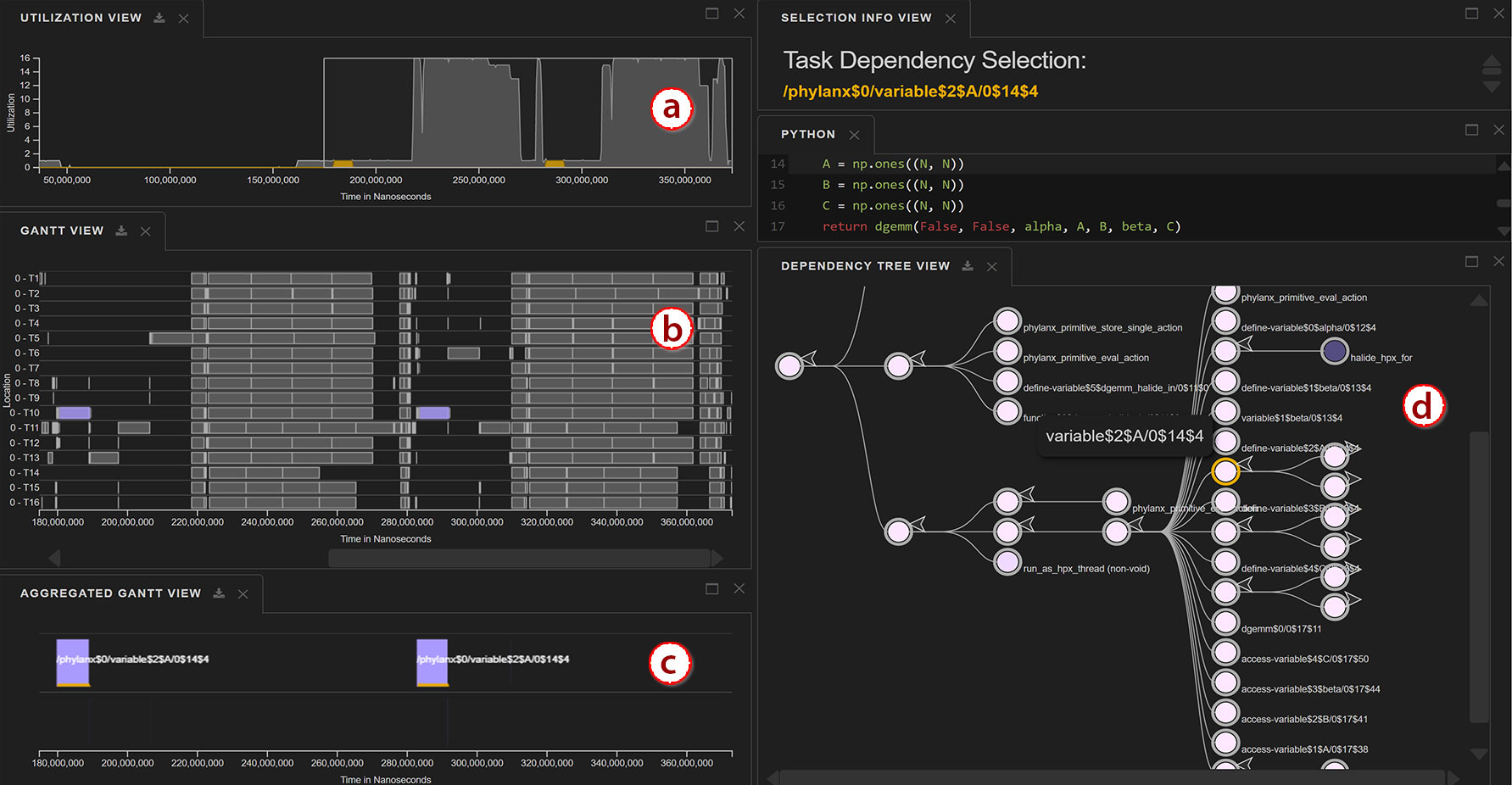}
    \hspace{1.5ex}
    \includegraphics[width=0.49\textwidth]{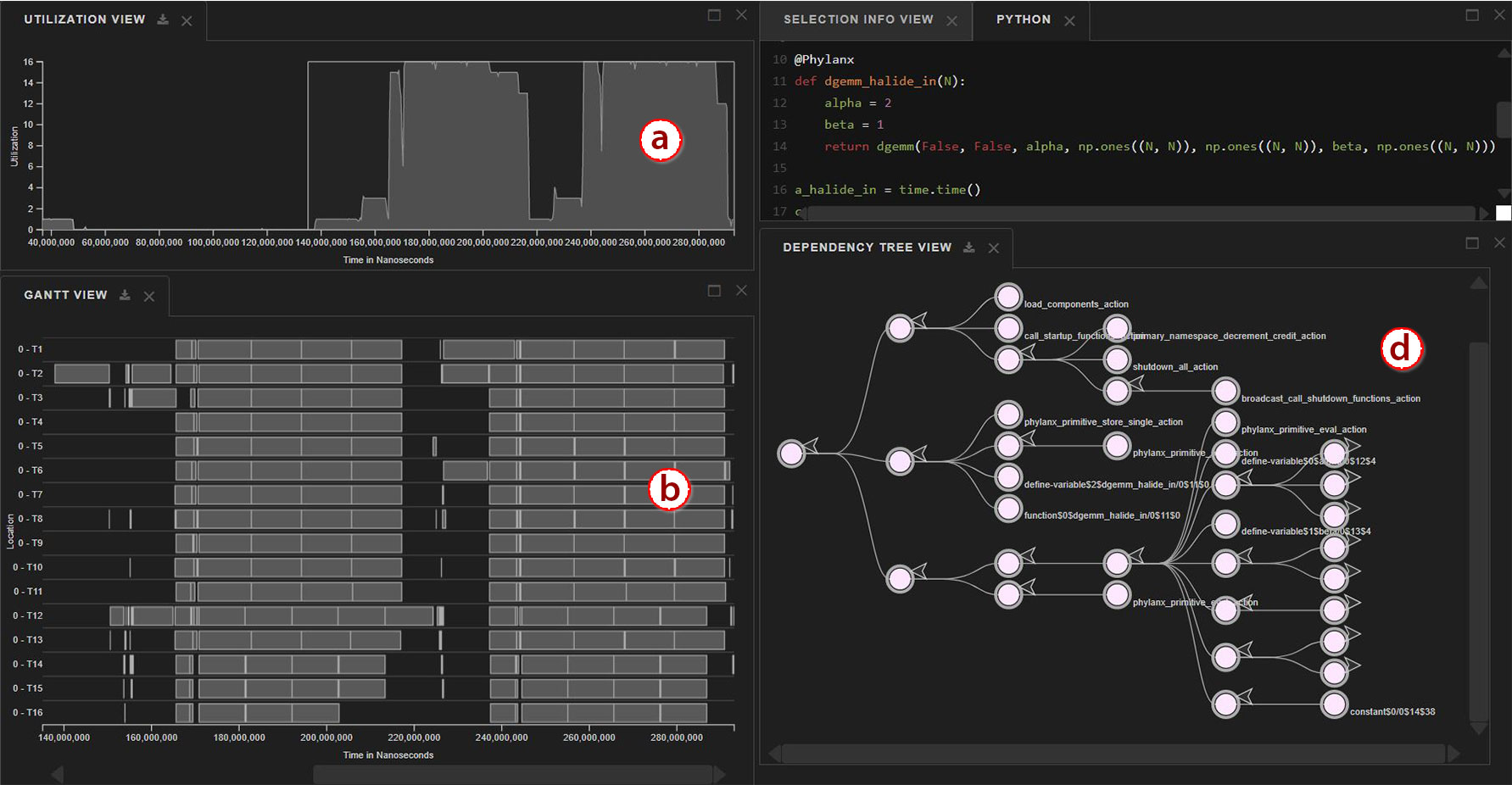}
    \caption{Traveler helped us identify a performance bug related to serialization of variable instantiation (left) leading to a fix that shortens execution time (right) and was merged into the Phylanx repository. On the left, we select a primitive $A$ in the Dependency Tree View (d) that it is executing alone as show in all temporal views (a, b, c). After we identified and fixed this bug, we are able to create multiple variables concurrently (right).
    }
    \label{fig:casestudy3}
\end{figure*}

During the previous analyses, we noted the long low-utilization periods between the two high-utilization phases of the program in the Utilization View (Goal G5). We used the Dependency Tree View to understand the mapping between primitives near the root (high level operations) and utilization (Sub-goal CC.3). As shown in \autoref{fig:casestudy3}(left), some of the low utilization can be attributed solely to the declaration of the variable $A$, as seen with the yellow area in the utilization view. We repeated the process with neighboring nodes for variables $B$ and $C$---each was the only active primitive in a separate area of low utilization. {\em Insight:} The declarations are serialized.

In the source code view, we saw that each declaration was of the form \texttt{variable = np.ones((N, N))}. {\em Insight:} The declarations could be executed concurrently. The RTM was not sure why this was happening, so we brought it to the attention of the Runtime Team lead. The Runtime lead explained it was a performance error because the way in which the {\tt dgemm} was implemented did not indicate the operations could be parallelized safely. He suggested a way that takes advantage of Phylanx's expression tree parallelization.  

After updating the code, we regenerated the OTF2 data, resulting in \autoref{fig:casestudy3}(right). The smaller gap between the two peaks in the utilization view verifies that 
the declaration of the variables $A$, $B$, and $C$ occurred in parallel. This bug fix was merged into the Phylanx repository.

\subsection{Feedback from Runtime Team}

We discuss feedback collected from Runtime Team members. First, we interview a former Runtime Team member who used Traveler for several months. Then, we introduce Traveler to novice Runtime Team members and gather their feedback. 

\vspace{1ex}

\noindent
\textbf{Feedback from an Expert User.}
One Runtime Team member (R1) used Traveler over a period of seven months. R1 communicated questions and reported bugs via e-mails. After they left the Runtime Team, we conducted a semi-structured interview with R1 regarding their experience. We summarize R1's feedback and associate their goals with our task analysis (\autoref{sec:taskanalysis}).

R1 said they switched from Vampir~\cite{nagel1996vampir} to Traveler because the latter is free. They noted {\em ``it was as satisfying as Vampir''}. R1 described their typical process as annotating their code, collecting a trace with the annotations, and then checking in Traveler to verify the annotated intervals were scheduled as their annotations expected. This analysis relates to sub-goal G5.1, correlating intervals with resource use. R1 frequently used the Gantt View, the Utilization View, and the Selection Info View in this endeavor. They said they primarily navigated via scrolling as their traces were not large.

Later, they switched from the program comprehension and verification concern to a performance analysis one. For their performance analysis work, they said they focused on the distribution of intervals (``tasks'') on resources (``cores'') and the dependencies between them: {\em ``I think we were mostly wanting to see how tasks are distributed among the cores and how they are being, how are the dependencies, how another core takes that part, what gets continued on another core''}. This is another example of G5.1, along with G4 in terms of following dependencies set by the runtime. 

We asked about the Interval Histogram and the Dependency Tree Views as other support for their goals. However, they noted these features were not yet available while they were using Traveler.

We asked if they also used VTune for their general performance analysis, as we are aware of other Runtime Team members who use it. They explained once they figured out Traveler, they mostly used it, noting, {\em ``Traveler was giving us everything we needed.''}

We asked about issues in Traveler. R1 discussed the bugs discovered during the development process, which they had previously e-mailed us about, and our in-development tagging system for managing datasets.

\vspace{1ex}

\noindent
\textbf{Novice Team Member Sessions.} We conducted feedback sessions with four novice Runtime Team members (P1-P4) individually. Each session was one hour long and conducted over videoconferencing. The session started with a briefing, followed by a demonstration with the {\tt dgemm} dataset. Participants were then asked to share their screen and attempt a series of tasks with a new dataset collected from an execution of the {\tt kmeans} algorithm. We then asked the participants about their experience and then debriefed them.

Prior to the session, the participants were provided with Traveler documentation, a video tutorial, and a link to a live demonstration. Only P4 said they viewed these materials. 

We designed the following tasks to exercise the goals from the task analysis, as annotated in parentheses. Each task could be completed in multiple ways. Participants started in the base Traveler configuration with only the Utilization, Gantt, Selection Info, and Source Code views. We asked the participants to load the Functional Box Plot View for E5:
\begin{enumerate}[font={\bfseries},label={E\arabic*.}]
    \itemsep=0ex
    \item Take 5 minutes to explore the dataset. What do you notice? (G1)
    \item Locate the longest interval bar. (G3, G4)
    \item Find the primitive with highest occurrence, in terms of total execution time. (G3)
    \item Locate highest CPU utilization and find the responsible primitive(s). (G2, G5)
    \item Explore the relationship between utilization and L1/L2 data cache misses. (G3, G4, G5)
\end{enumerate}

We then conducted a semi-structured interview with the participants. The following questions were asked during the interview:
\begin{enumerate}[font={\bfseries},label={Q\arabic*.}]
    \itemsep=0ex
    \item Which view helped the most to perform the tasks?
    \item which feature did you like most?
    \item What other tools you have used and how do they compare?
    \item Are there additional features which we could include?
    \item How long have you been working with the HPX team?
    \item Do you have any other comments?
\end{enumerate}

P1 had no prior experience with HPX, P2 had nine months, and P3 had ``a few'' months. P4 had been with the team for two years, but not worked with HPX for the entire time. P1-P3 had no experience using a performance visualization tool. P4 had experience with VTune.

During the task phase, P1 and P3 struggled with the system as a whole, including operations such as adjusting tabs and panning the view. They did not complete any tasks except finding the highest utilization, so we omit discussing them further. The transcripts for their sessions along with the others are available in the supplemental materials.

In E1, P2 explored most of the views, starting by reading the source code, explaining it, and then using the Dependency Tree View and Utilization View to recognize high utilization due to a primitive {\tt k\_means\_t}. As P2 was familiar with HPX but not Phylanx, they asked questions about what some leaf nodes in the tree did. P4 took a different strategy of browsing the Gantt View via zooming. We asked why they did not use the Utilization View to zoom faster---they did not notice the brush which was scaled to the full chart at first. P4 noted the intervals were much shorter in comparison to the tutorial trace.

P2 completed E2, finding the longest interval, quickly with the Interval Histogram View. P4 tried browsing the Gantt View but was confused by the illusion of many tiny intervals as one large bar. With a suggestion from the facilitator, they completed the task.

In E3, finding the primtive with the highest occurrence, P2 looked at the Interval Histogram View before browsing the Dependency Tree View and determining the answer. P4 browsed the Dependency Tree View but did not suggest an answer. When we explained the result, P4 said he had not understood what was being asked and had actually noticed the node of interest fairly early.

E4 had two sub-trials---locating highest CPU utilization and finding out the responsible primitive for the highest CPU utilization. Both P2 and P4 were able to quickly determine the highest utilization and then successfully determined the source using the Dependency Tree View and trace-back dependencies in the Gantt View, although some hints were provided to P4 in the second part. 

In E5 participants were asked to explain data cache misses in association with CPU utilization. P2 discerned the relationship between the level 1 data cache miss and how CPU utilization affected that. P4 also identified the fluctuations in level 1 data cache misses. Neither participant commented on level 2 data cache miss.

\vspace{1ex}

\noindent\textbf{Novice Feedback Discussion.} P2 was able to complete all tasks. In E1, the overview (G1) task, they explained the program and went a step further with the insight of identifying high utilization (G5) and attribute it to a primitive (CC.3), preemptively performing task E4. They accomplished the extreme interval (E2: G3.6) and correlation of L1 cache and utilization (E5: CC.4, G5) quickly. The latter led to the insight that higher utilization led to more cache misses. E3, finding the primitive responsible for the most time (CC.3) took more time as they first selected the view showing distribution of lengths rather than aggregated time. P4 was able to complete most tasks but needed reminders despite being the one who had consulted the materials beforehand. While P2's results were encouraging, we acknowledge these are limited results.

During the interview, both P2 and P4 said the Dependency Tree View was useful. P2 also spoke about the Utilization and Functional Box Plot Views, with the Functional Box Plot being their favorite. P4 also mentioned the Gantt View, saying they needed both that and the Dependency Tree View for their work. The linking between multiple views was their most-liked feature.

In terms of requested features, P2 suggested a multi-select for interval bars in the Gantt View. P4 suggested coloring the interval bars by a metric, such as the performance counters.

We noticed some usability issues. Participants were less familiar with brushing, wanting to click on the Interval Histogram View bars or not noticing the brush in the Utilization View. Also, when the Gantt View has many small intervals, it draws them with the outline color (dark gray) only. Familiar users can distinguish, but new users thought they were seeing one large bar, suggesting other signifiers are needed when intervals are dense.

\vspace{1ex}

\noindent
\textbf{Local versus Cloud Performance.} In the evaluation sessions, we noticed additional latency in updating the Gantt View. We profiled a variety of interactions on the \texttt{kmeans} test dataset (29K trace events) deployed locally and online for the study. Locally, Gantt View updates took between 446ms (single click selection) and 3,643ms (zooming). In the cloud deployment, the interaction extents were between 907ms and 4,018ms respectively. Both versions took between 63 and 1,863ms for client-side rendering, but the cloud version experienced a 56.8\% increase in network transfer and resource loading times. Some of the difference may also be due to the limitations of the cloud service tier used, which offers 500MB of RAM compared to the 16GB in the local deployment. More details regarding this preliminary performance analysis are in the supplemental materials.

Even in the local case with this relatively small dataset, some updates could take seconds, though not all zoom interactions took that long. Further performance analysis and optimization is needed to improve interactive scalability, including across other views and with focus on both client and server side operations.

\subsection{Limitations and Threats to Validity}

In the case study, the Visualization Team member was the primary user of Traveler for the findings shown, though the Runtime Team member had used it for overviews previously. The case study resulted in new insights, but did not exercise the usability of Traveler directly with a member of the target user group.

The interview with the expert user R1 was over three months since they had last used Traveler and said they did not remember some details. Also, R1's case did not require a large dataset. Their use of Traveler was before all of the views presented here were completed, so we do not have deployment feedback related to those newer views.

We collected feedback from only five people, of which only one was a true expert. Thus, our results are preliminary and limited. Furthermore, 
as we are collaborators with the Runtime Team, and though we had not met the novice users before, all users may have been biased towards positive feedback. 

Many of the novice users had been working with the Runtime Team for a year or less and thus had very little experience. We had piloted the evaluation session with one expert and one person with passing knowledge, because at the time we did realize how new the team members would be compared to previous sessions we had done with the Runtime team. Thus, our explanations may not have been appropriate to the participants' experience levels.

The cloud deployment of Traveler incurred more interaction latency than the local version, which may have affected the participants' analysis process. However, both versions can sometimes experience several second delays with the given datasets, mitigating the differences.

\section{Reflections and Lessons Learned}
\label{sec:reflection}
We reflect on the design of Traveler, in particular on (a) our observations regarding a new visualization in an ongoing collaboration with rapidly changing concerns as well as (b) our experience with multi-scale, configurable designs for execution trace data.

\vspace{1ex}

\noindent\textbf{We were able to leverage an existing task analysis, but there were challenges as the data significantly changed.} The task analysis presented in \autoref{sec:taskanalysis} is a revision of a task analysis performed earlier in the collaboration when we were focusing on execution tree profile data---data that is aggregated by primitive context. As we had before, we continued to track tasks over our meetings, reinforcing tasks that repeated over time while de-prioritizing tasks that appeared infrequently. The higher levels of the task analysis, which we referred to as umbrella concerns and goals, were mostly unchanged, while sub-goals proliferated. The relative stability of goals gave us confidence in our prioritization of features, especially ones where the new data allowed us to expand our support for an existing goal or sub-goal. Furthermore, as the many of the goals behind the earlier work were maintained over time, we were able to transfer parts of the previous visualization design as a view in Traveler, again with a prioritization of which design features should be ported.

However, we ran into difficulties at the abstract visualization task level because the primary data abstraction changed from the execution tree graph to a parallel execution trace. With the execution tree, we had applied the network task taxonomy of Lee et al.~\cite{Lee2006}, which no longer fully described our data. We chose to use another task taxonomy~\cite{Yi2007} in tandem and describe needs to our best effort. We did not find this process as helpful as in the previous project, where the tasks so clearly pointed to a visual idiom. We question how much richness was added over our sub-goals. These experiences may indicate that either the sub-goals are enough, we need to further sub-divide sub-goals, or we need to find a better fit in describing the performance analysis process.

\vspace{1ex}

\noindent\textbf{There is much more opportunity to design across experience levels, but it is unclear how to prioritize.} The newest Runtime Team members struggled to use Traveler. Those with a little more experience completed most tasks, though some with help. Our domain expert user said they were able to switch from a commercial trace visualization tool seamlessly but did not explore newer views. Integrating some form of training into Traveler may help all of those classes of users.

We offered documentation with example analyses, images, and a movie, but the one participant who engaged with them still had difficulties. It is unclear how much training is necessary and how it can be done in a scalable manner.

\vspace{1ex}

\noindent\textbf{Immersive practices~\cite{Hall2019} during the case study helped refine the tool.} Early Traveler development used several data sets to test features and visualization performance, but there were not analysis goals associated with them. The case study provided a real analysis goal. By driving the visualization through the case study, the Visualization Team identified new bugs and performance bottlenecks in Traveler as well as workflow and interaction issues.

\vspace{1ex}

\noindent\textbf{It is beneficial to automate techniques for deployment as soon as possible, but unfortunately not always practical.} As this design study is a collaboration with a rapidly developed open-source project, meaningful data, e.g., data representing the latest state of the runtime, was constantly changing. Automating the pipeline from library updates to application updates to new trace data representing those updates to Traveler's pre-processing and ultimately the visualization was a boon to keeping the Visualization Team in step. However, having automated our own pipeline, we overlooked the pipelines of our collaborators. Runtime Team members were often unaware of new features and did not fetch the latest Traveler version. 

\vspace{1ex}

\noindent\textbf{Multiple levels of abstraction between overview and detail helped us navigate the trace.} The scale difference between the entire execution length of the trace, represented by the Utilization View, and individual intervals, shown in the Gantt View, is vast. We implemented multiple levels of abstraction in between, leveraging the known structures of the source code, which organizes by line, and the individual primitives within a line of source code, as accessed through the Dependency Tree View and Aggregated Gantt View. We used these intermediary abstractions to discover the variable instantiation performance bug in our case study.

\vspace{1ex}

\noindent\textbf{Participants made much use of the configurable views. The overview was the only constant.} During our evaluation sessions, we observed that each participant rearranged and resized views throughout the tasks.  This behavior is resonant as Traveler has many views, some of which support fewer sub-goals than others, so users configuring the views for a sub-goal of focus was intended by the design. 

We did not notice trends in their arrangements, but our number of participants was small. We did notice one constant in their choice of views---the Utilization View was always present. It served as a major navigational aid for temporal views as well as represented one of the high level goals of performance analysis, G5, understanding utilization.

\section{Conclusion}
\label{sec:conclusion}
We presented the design of Traveler, a highly linked multi-view coordinated visualization for visual exploration of task parallel execution traces. To manage the vast scale differences in the trace data and the irregular nature of asynchronous task scheduling in a way users could relate to their code, we introduced linked execution tree and aggregated Gantt views. We further supported analysis tasks with through linked performance counter summaries and an interval histogram. In our case study, we were able to explain performance behavior of a runtime and found a performance bug that was ultimately fixed and committed to our collaborator's repository. In addition to our visualization design and strategy, through this design study we contribute further insights about ongoing design collaborations, adapting task analyses and considerations for the changing team.

 \acknowledgments{This work was supported by the United States Department of Defense through DTIC Contract FA8075-14-D-002-007, the Department of Energy under DE-SC0022044, and the National Science Foundation under NSF IIS-1844573.}

\bibliographystyle{abbrv-doi}

\bibliography{main}
\end{document}


\firstsection{Guide to Supplemental Materials}

\maketitle

This document contains description of all supplemental material files (below). We then provide more details regarding the Traveler system design, including links to the repository and live demo. We follow this with details about the case study settings and figures that we could not fit in the main document. Finally, we provide details regarding the procedure for our preliminary quantitative performance analysis for the Gantt View.

\section{Included Files}
All materials are included in the accompanying {\em .zip} file as:

\begin{itemize}
    \itemsep=0ex
    \item {\bf TravelerDemonstration.mp4 (Length 6 minute 28 seconds)} - Video demonstration of available features offered by Traveler for performance analysis of task parallel traces.
    \item {\bf TaskAnalysisUpdateAnonymized.md} - Task analysis.
    \item {\bf CaseStudyAnalysis.md} - Case Study Summary Presented to the Runtime Team.
    \item {\bf cache\_misses.JPG} - Cache miss reported to the Runtime team.
    \item {\bf CaseStudyFeedbackEmail\_1.txt} - First feedback emails on the case study by the runtime team.
    \item {\bf CaseStudyFeedbackEmail\_2.txt} - Second feedback emails on the case study by the runtime team.
    \item {\bf ContinuousDeploymentEmails.txt} - Emails during continuous deployments.
    \item {\bf interview\_script\_R1.txt} - Interview script for collecting feedback from non-authoer member (R1) of runtime team.
    \item {\bf recording\_transcript\_R1.txt} - Interview Recording transcript of R1.
    \item {\bf transcript\_R1\_coded.txt} - Coded interview recording transcript of R1.
    \item {\bf R1\_code\_summary.md} - Merged codes and summary of interview recording transcript of R1.
    \item {\bf interview\_script\_2.txt} - Interview script for collecting feedback from novice team members of runtime team.
    \item {\bf TravelerIntro.md} - An introductory documentation of Traveler describing all available features.
    \item {\bf TravelerSampleTasks.md} - A step by step walk-through with screenshots of some sample tasks in Traveler.
    \item {\bf interview\_recording\_transcript\_P1.txt} - Interview recording transcript of P1.
    \item {\bf interview\_recording\_transcript\_P1\_coded.txt} - Coded interview recording transcript of P1.
    \item {\bf interview\_recording\_transcript\_P1\_summary.md} - Merged codes and summary of interview recording transcript of P1.
    \item {\bf interview\_recording\_transcript\_P2.txt} - Interview recording transcript of P2.
    \item {\bf interview\_recording\_transcript\_P2\_coded.txt} - Coded interview recording transcript of P2.
    \item {\bf interview\_recording\_transcript\_P2\_summary.md} - Merged codes and summary of interview recording transcript of P2.
    \item {\bf interview\_recording\_transcript\_P3.txt} - Interview recording transcript of P3.
    \item {\bf interview\_recording\_transcript\_P3\_coded.txt} - Coded interview recording transcript of P3.
    \item {\bf interview\_recording\_transcript\_P3\_summary.md} - Merged codes and summary of interview recording transcript of P3.
    \item {\bf interview\_recording\_transcript\_P4.txt} - Interview recording transcript of P4.
    \item {\bf interview\_recording\_transcript\_P4\_coded.txt} - Coded interview recording transcript of P4.
    \item {\bf interview\_recording\_transcript\_P4\_summary.md} - Merged codes and summary of interview recording transcript of P4.
    \item {\bf traveler\_qeval.xlsx} - Quantitative analysis spreadsheet.
    \item {\bf QuantitativeEval.zip} - Inspection records of Google Chrome during quantitative analysis.
\end{itemize}

\section{Traveler}
\label{sec:traveler}

\noindent
Traveler live demo link (herokuapp.com): \href{https://traveler-integrated.herokuapp.com/}{Traveler}

\noindent
Traveler Github location link: \href{https://github.com/hdc-arizona/traveler-integrated}{Traveler-git}

\vspace{1ex}

\begin{figure}[h]
  \centering
  \includegraphics[width=\linewidth]{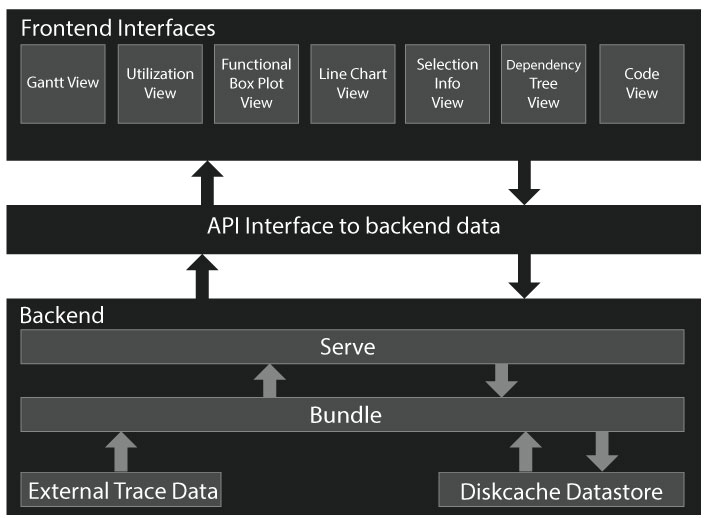}
  \caption{Traveler System Architecture. The Frontend interface contains the interactive views. The backend reads, stores, and preapre trace data for the frontend interface. An intermediate API interface is used to supply data between the backend and frontend interface. }
  \label{fig:tarch}
 \end{figure}
 
\noindent
{\bf Traveler System Architecture}: 
Traveler system architecture is provided in \autoref{fig:tarch}. The Traveler follows MVC (model-view-controller) design pattern for developing the user interface, partitioning programming logic, and backend data handling. While logic and components for the frontend (view and controller) are written on javascript and presented on an interactive web-interface, the logic for backend data structure, handling, parsing, storing, querying are written on python.

The libraries and programming languages used in the backend and fronted are summarize in \autoref{tab:back_libs} and  \autoref{tab:front_libs} respectively.

\begin{table}
    \centering
    \caption{Libraries and languages used in the backend}
    \label{tab:back_libs}
    \scriptsize
    \begin{tabular}{|l|}
        \hline
        newick $\ge$ 1.0.0, \\
        sortedcontainers $\ge$ 2.4.0, \\
        aiofiles $\ge$ 0.5.0, \\
        fastapi $==$ 0.65.2, \\
        uvicorn $==$ 0.11.7, \\
        python-multipart $\ge$ 0.0.5, \\
        diskcache $\ge$ 4.1.0, \\
        numpy $\ge$ 1.19.1, \\
        cffi $\ge$ 1.14.1, \\
        interval tree - [Redacted] \\
        \hline
    \end{tabular}
\end{table}

\begin{table}
    \centering
    \caption{Libraries and languages used in the frontend}
    \label{tab:front_libs}
    \scriptsize
    \begin{tabular}{|l|}
        \hline
        ukijs/ui $\ge$ 0.2.5, \\
        codemirror $\ge$ 5.60.0, \\
        d3 $\ge$ 6.6.1, \\
        golden-layout $\ge$ 1.5.9, \\
        jquery $\ge$ 3.6.0, \\
        less $\ge$ 4.1.1, \\
        oboe $\ge$ 2.1.5, \\
        uki $\ge$ 0.7.9 \\
        \hline
    \end{tabular}
\end{table}

\begin{table}[h]
    \centering
    \caption{Rostam Cluster Specification}
    \label{tab:rspec}
    \scriptsize
    \begin{tabular}{|l|l|}
        \hline
        CPU Model & Intel(R) Xeon(R) Gold 5118 \\
        \hline
        CUU Clock Speed & 2.30GHz \\
        \hline
        Number of Cores & 24 \\
        \hline
        RAM & 96GB \\
        \hline
    \end{tabular}
\end{table}

\begin{table}
    \centering
    \caption{Library versions}
    \label{tab:libs}
    \scriptsize
    \begin{tabular}{|c|c|}
        \hline
        library & Commit \\
        \hline
        HPX & \href{https://github.com/STEllAR-GROUP/hpx/tree/dc79a36c922087edf2653fae5d0f2dd72706ab3b}{dc79a36c92} \\
        \hline
        Phylanx & \href{https://github.com/STEllAR-GROUP/phylanx/tree/295b5f82cc39925a0d53e77ba3b6d02a65204535}{295b5f82cc} \\
        \hline
        Halide & \href{https://github.com/halide/Halide/tree/5dabcaa9effca1067f907f6c8ea212f3d2b1d99a}{Version 12.0.1} \\
        \hline
        Phylanx-Halide & \href{https://github.com/STEllAR-GROUP/phylanx_halide/tree/12ab94bc3a93efcf63b088a125c999b192d36e18}{12ab94bc3a} \\
        \hline
        Blaze & \href{https://bitbucket.org/blaze-lib/blaze/commits/c4d9e85414370e880e5e79c86e3c8d4d38dcde7a}{c4d9e85414} \\
        \hline
        Blaze Tensor & \href{https://github.com/STEllAR-GROUP/blaze_tensor/tree/4b6ec2c6b20b46de36ac4374deea03d2fc4b3753}{4b6ec2c6b2} \\
        \hline
        Traveler & \href{https://github.com/hdc-arizona/traveler-integrated/tree/ea668d7349157a56c3585abe10418b338f850823}{ea668d7349} \\
        \hline
    \end{tabular}
\end{table}

\section{Case Study}
\label{sec:case}

\begin{figure}[!h]
    \centering
    \includegraphics[width=0.49\textwidth]{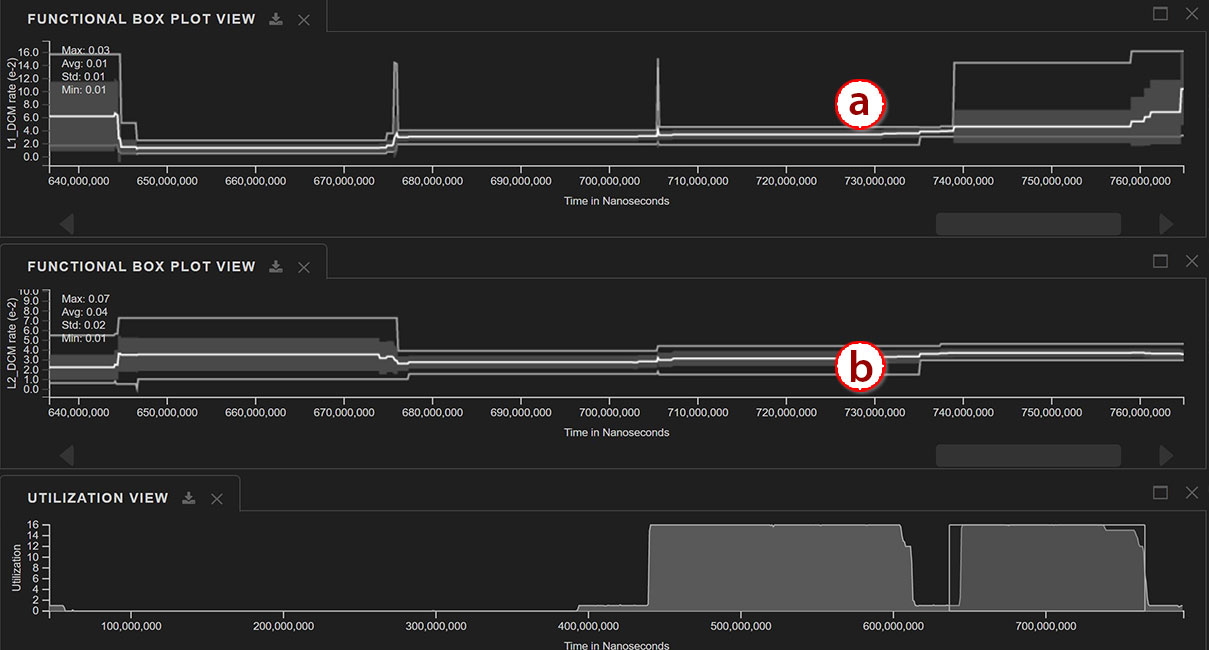}
    
    \vspace{1.5ex}
    
    \includegraphics[width=0.49\textwidth]{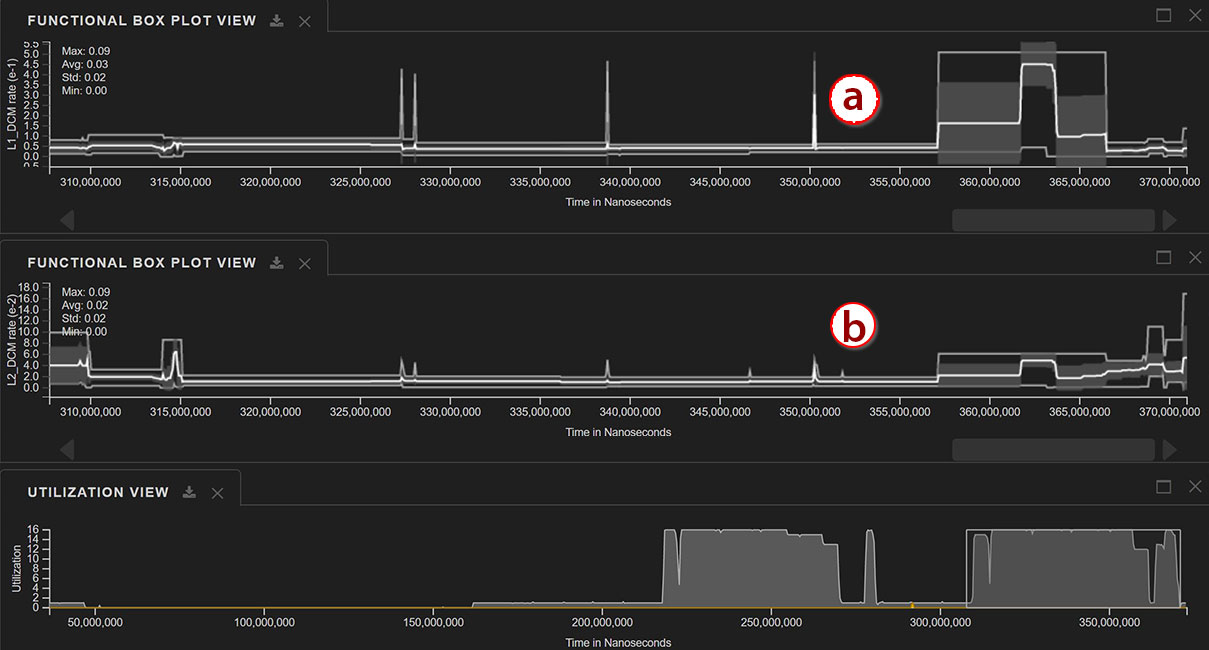}
    \caption{Traveler helps identify cache behavior differences between Phylanx-Blaze (top) and Phylanx-Halide (bottom). The Functional Box Plots of L1 (a) and L2 (b) cache miss rates show the L2 rate rises steadily during the execution in Phylanx-Blaze while the Phylanx-Halide version remains flat.  
    }
    \label{fig:casestudy2}
\end{figure}

\autoref{fig:casestudy2} shows the differences between cache behavior that Traveler helped us identify. 

All relevant experiments for the case study was conducted on the Rostam cluster at the Center for Computation and Technology at Louisiana State University {(LSU)}. The platform specification is summarized in~\autoref{tab:rspec}. The libraries used during the case study have been summarized in~\autoref{tab:libs}.

\section{Gantt View Performance Analysis}
\label{subsec:quanteval}
We conduct a quantitative profiling of Traveler by estimating the interaction time, the time difference between user's interaction with the interface (mouse click, scrolling, panning) and interface update finish time. We utilized the performance profiling tool of Google Chrome to conduct the quantitative analysis. We execute our experiment while running the Traveler server both locally and on a Cloud platform. We used Google Chrome client browser to browse the Traveler interface. Then using the inspect feature, we recorded our interactions and conducted profiling. The local platform specification is,
\begin{itemize}
    \item CPU: Intel Core i7-8750H CPU @ 2.20GHz
    \item RAM: 16GB
    \item OS: Windows 10
\end{itemize}
During the analysis, the remote platform specification is, 
\begin{itemize}
    \item Cloud Platform: Free Heroku Dyno.
    \item RAM: 512MB.
    \item OS: Ubuntu 20.04
\end{itemize}
The recorded profiled data is in {\em QuantitativeEval.zip} and summarized interactions over the Gantt View is in {\em traveler\_qeval.xlsx}.

\bibliography{main}